\documentclass[sigconf]{acmart}
\AtBeginDocument{%
  }

\setcopyright{acmlicensed}
\copyrightyear{2026}
\acmYear{2026}
\setcopyright{cc}
\setcctype{by}
\acmConference[KDD '26]{Proceedings of the 32nd ACM SIGKDD Conference on Knowledge Discovery and Data Mining V.2}{August 09--13, 2026}{Jeju Island, Republic of Korea}
\acmBooktitle{Proceedings of the 32nd ACM SIGKDD Conference on Knowledge Discovery and Data Mining V.2 (KDD '26), August 09--13, 2026, Jeju Island, Republic of Korea}
\acmISBN{979-8-4007-2259-2/2026/08}
\acmDOI{10.1145/3770855.3818648}

\acmSubmissionID{v2bsi058}

\usepackage{subcaption}

\begin{document}

\title{Towards Auditing AI Systems in the Wild}

\author{Aditya T. Vadlamani}
\orcid{0009-0003-8713-2566}
\email{vadlamani.12@osu.edu}
\affiliation{%
  \institution{Ohio State University}
  \city{Columbus}
  \state{Ohio}
  \country{USA}
}

\author{Anutam Srinivasan}
\orcid{0009-0000-7982-7251}
\email{asrinivasan350@gatech.edu}
\affiliation{%
  \institution{Georgia Institute of Technology}
  \city{Atlanta}
  \state{Georgia}
  \country{USA}}

\author{Srinivasan Parthasarathy}
\orcid{0000-0002-6062-6449}
\email{srini@cse.ohio-state.edu}
\affiliation{%
  \institution{Ohio State University}
  \city{Columbus}
  \state{Ohio}
  \country{USA}
}

\begin{abstract}
AI systems are increasingly deployed in real-world settings where their behavior is shaped by dynamic environments, evolving data distributions, and complex interactions with users and infrastructure. Traditional machine learning evaluation focuses on benchmarks and operates within sandboxed environments, providing only a limited view of the true system behavior \textit{in the wild}. We argue for the development of principled auditing frameworks that monitor deployed AI systems throughout their lifecycle. We further propose framing auditing as a statistical problem of monitoring constraint violations under uncertainty, where desired properties (e.g., fairness and safety) are treated as risk-controlled constraints that must be continuously evaluated as systems evolve through iterative feedback. This perspective highlights the need for uncertainty-aware monitoring methods, socio-technical specifications of audit criteria, and auditing infrastructures that enable ongoing oversight of AI systems in the wild.
\end{abstract}



\maketitle

\section{Auditing AI: A Clear and Present Need}
Recent advances in machine learning have led to rapid development and deployment of AI systems across many real-world domains. Today, AI and ML systems shape information access~\citep{lecun2015deep} and decision-making in areas ranging from recommendation systems~\citep{aggarwal2016recommender} and financial services~\citep{dixon2020machine,wuthrich2023statistical} to AI assistants in healthcare~\citep{Bhuyan2025-zu}, autonomous agents~\citep{park2023generative}, and embodied agents~\citep{feng2025embodied}. As these systems are increasingly embedded in social and institutional processes, ensuring their trustworthiness has become a core challenge. 

Despite the growing societal impact, the dominant evaluation paradigm in machine learning is mainly \textit{offline} and \textit{pre-deployment}.
In practice, many of the most consequential behaviors of AI systems only emerge after deployment.  
These challenges are amplified by the rapid pace of innovation in AI \cite{bengio2026international}. Organizations face strong incentives to deploy increasingly capable models quickly, with the ever-evolving landscape and competition between frontier AI players (e.g., DeepMind~\citep{team2023gemini}, Anthropic~\citep{anthropic2025system}, \& OpenAI~\citep{singh2025openai}). This dynamic creates a ``race'' in which organizations prioritize immediate deployment over careful evaluation and long-term risk assessment. Organizations that largely ignore or pay lip service to safety precautions may win the race in the short term, but at a cost to society in the long term.

Researchers have highlighted the vulnerabilities in deployed AI systems, including risks related to safety, legality, and security, as well as their potential for discriminatory effects~\citep{mokander2023auditing, hirsch2024business}. As AI systems continue to expand in capability and influence, public trust in such systems is at a tipping point. Mechanisms that enable independent verification of claims about system performance, fairness, and reliability are increasingly important \citep{maneriker2023}. Effective auditing of AI systems offers a promising approach to addressing this challenge by assessing an organization's claims about algorithmic efficacy, safety, and security, while helping organizations detect failures, mitigate fraud, and reduce discrimination~\citep{gabriel2024ethics}. Building such trust is the essential fabric of modern society~\citep{arrow1974limits}. 

Effective and efficient auditing mechanisms enable the careful deployment of AI systems and their democratic governance~\citep{falco2021governing}. In short, auditing is essential for achieving trustworthy AI, but the challenges are daunting.  First, the proprietary nature of many AI systems makes auditing them challenging. Second, auditing requires a quantifiable specification (from a legal, governance, safety, and security standpoint, grounded in societal norms and specific silos (e.g., health, finance)). Third, the socio-technical ability to assess whether this specification continues to hold once such systems are deployed at scale in the wild is non-trivial. Fourth, audits may be public (black-box audits) or private (grey- and white-box audits conducted by trusted third parties). Relatedly, audits are required not only for the final product but also for the development process. Finally, to ensure efficient deployment, audits should be seamless, fast, secure, and accurate--enabling organizations to serve their clients and/or the public effectively.

\section{Dimensions of AI Auditing}
Modern AI deployments involve complex interactions among models, data pipelines, infrastructure, and human users, and as a result, auditing cannot be treated as a single evaluation task; it must instead consider several complementary dimensions.

\subsection{Lifecycle}
\subsubsection{Pre-Deployment AI Evaluations}
The dominant paradigm for evaluating AI systems has primarily focused on \textbf{pre-deployment evaluations}, where models are trained on historical data and evaluated on benchmark datasets or held-out test sets prior to deployment. This paradigm enabled rapid progress in algorithmic development and comparisons of model performance, further motivating innovations by organizations and research labs. However, as AI systems are increasingly deployed in dynamic real-world environments, benchmarking often provides only a narrow view of system behavior~\citep{raji2021AI}. Importantly, such benchmarks can be manipulated and engineered (see, for example, Pendragon's case against Sun Microsystems on the CaffeineMark Benchmark in the 1990s~\citep{java1997} or the more recent Volkswagen ``dieselgate'' scandal). In practice, the reliability, safety, and societal impact of AI systems are determined not only by their performance on static benchmarks but also by their interactions with evolving environments, infrastructure, and human users. Several features of modern AI systems limit the effectiveness of a purely pre-deployment evaluation.

\paragraph{}\hspace{-5.5mm}{\textbf{Limited Visibility into Complex AI Systems.}}
Modern AI systems rarely consist of a single model operating in isolation, but instead include several models embedded within complex pipelines that include data collection, model training, deployment infrastructure, and downstream decision-making processes~\citep{sculley2015hidden}. These systems, in turn, interact with external components (e.g., data sources, other automated systems), raising questions about the \textit{provenance} of AI systems. As a result, evaluating a model on a benchmark dataset provides a limited view of the broader system behavior. Furthermore, AI system internals and decisions made during development are not well understood (or even shared), resulting in sources of uncertainty that are ignored by evaluation or auditing. For example, a key component of model development that is commonly overlooked concerns how data is collected (i.e., data provenance) for training, as well as whether data from diverse contexts (e.g., different socio-cultural groups) is well-represented. Finally, system failures may encompass issues in data collection or system integration, highlighting the incompleteness of existing model evaluation mechanisms for assessing the \textit{entire} AI system pipeline.

\paragraph{}\hspace{-5.5mm}{\textbf{Real-World Dynamics.}}
AI systems that are deployed in real-world settings must operate in inherently dynamic environments--data distributions may shift over time~\citep{Al_Maliki_2024}, user behavior may evolve in response to system outputs~\citep{Chaney_2018}, and new or unintended use cases may emerge after deployment (see, for example, Microsoft's Tay chatbot from 2016). These dynamics can produce behaviors that are difficult to anticipate during development. For example, model performance can degrade due to distribution shift~\citep{liu2023outofdistributiongeneralizationsurvey} or amplify biases through human-AI feedback loops~\citep{Glickman2025-xh}. This is exacerbated for embodied AI, where distribution shifts can lead to unreliable performance and dangerous behaviors (i.e., obstacle collisions)~\citep{srinivasan2026safety}. Pre-deployment testing can help identify certain classes of failures, but it cannot fully anticipate how systems will behave in complex social and technical environments.

\paragraph{}\hspace{-5.5mm}{\textbf{Incentives for Rapid Deployment.}}
The aforementioned challenges are compounded due to the incentive to rapidly develop and deploy AI systems. Frontier AI companies operate in a competitive environment in which rapid innovation and deployment are required to remain relevant and deliver significant advantages. As a result, organizations compete by prioritizing incremental improvements and rapid release cycles to remain competitive. The emphasis shifts to achieving state-of-the-art performance on widely used \textit{existing} benchmarks and product metrics, rather than developing robust mechanisms for long-term understanding of behavior and risks.  While this perspective has led to remarkable AI capabilities, it disincentivizes a focus on comprehensive evaluation and oversight. The gap between technological advancement and mechanisms for monitoring, governing, and auditing continues to widen as we advance in the former while neglecting the latter.

\paragraph{}\hspace{-5.5mm}{\textbf{The Need for Auditing Across the AI Lifecycle.}}
Altogether, these factors highlight the limitations of evaluation frameworks that solely focus on model performance prior to deployment. Ensuring AI systems remain reliable and trustworthy requires mechanisms that extend beyond traditional benchmark-based evaluations and encompass the \textit{entire lifecycle} of AI systems. In particular, there is a need for approaches that enable \textbf{ongoing auditing of AI systems during development and deployment}, allowing researchers, organizations, and regulators to monitor system behavior, detect emerging risks, and intervene when necessary. Such approaches can help bridge the gap between static (model) evaluation and the dynamic environments in which the AI systems actually operate. This leads us to critically understand what \textbf{post-deployment evaluations (or auditing)} entails.

\subsubsection{Post-Deployment AI Auditing}
Auditing comprehensive AI systems after deployment poses additional challenges and questions. Incorrect usage of AI models can lead to catastrophic performance \citep{polyzotis2019data}, thereby flagging a model as unsafe post-deployment. However, the scope of the audit may be focused on compliance when the model is used \textit{within reason}. Thereby, introducing ambiguity, i.e., ``what is \textit{within reason}?'', into the auditing problem and requiring careful treatment in scoping \citep{kolt2026legal}. Furthermore, large-scale audits will require automated data filtration of data collected during deployment to align with the scope of the auditing problem, thereby increasing the scope for errors in the auditing pipeline. 

A well-known example of the complexity of post-deployment auditing arose in the case of the COMPAS recidivism risk assessment tool used in U.S. criminal justice systems. An investigation by ProPublica argued that the system exhibited racial disparities in false positive rates when predicting recidivism risk, suggesting discriminatory outcomes against Black defendants~\citep{angwin2016machine}. In response, the system’s developer contested these findings, arguing that the model satisfied an alternative fairness notion based on calibration across groups~\citep{dieterich2016compas}. The resulting debate highlighted that post-deployment auditing is not merely a technical task but also depends critically on the specification of auditing criteria and the interpretation of statistical evidence. Without clear specifications for fairness or risk thresholds, different auditing analyses may reach conflicting conclusions despite relying on the same underlying data.


Post-deployment auditing further encounters data heterogeneity, with the ubiquitous use of the same AI models, e.g., how educators use AI to create lesson plans, vastly differing from how software developers use them. Audits that focus on global performance across all tasks (e.g., toxicity in LLMs) will require careful handling to accurately reflect the model's behavior, even when certain tasks are prone to eliciting non-compliant behavior. 

Unlike pre-deployment auditing, correcting compliance issues in deployed models also poses several challenges: corrections must be made on the fly under expedited timelines while addressing the audit results. These constraints preclude full-model retraining and require us to consider approaches that seamlessly transition model audits to model updates.

\subsection{Access Levels and Institutional Roles}
Another important dimension of AI auditing concerns both the level of access that auditors have to the system and the institutional actors responsible for conducting the audit. In practice, these two aspects are closely related: different stakeholders typically operate under different levels of system visibility. For example, external researchers often rely on black-box interaction with deployed systems, whereas internal teams may have full (white-box) access to model internals and training pipelines.

As more and more AI companies compete to develop and deploy their own models, to maintain a competitive advantage, companies are becoming less transparent and more restrictive about their data, model architectures, and other system components, making black-box auditing techniques more desirable~\citep{maneriker2023}. That said, certain companies do (partially) open-source these details (e.g., Llama 4, Qwen, DeepSeek), resulting in wider adoption since end-users can independently evaluate model performance.

Table~\ref{tab:audit_roles} summarizes common auditing actors and the levels of system access typically available to them. These categories are not mutually exclusive, but they illustrate the range of possible auditing arrangements in practice and highlight the importance of a diverse auditing ecosystem. No single actor has complete visibility into the behavior and risks of complex AI systems. Internal teams possess the deepest technical access, while external researchers and users often observe real-world behaviors that may not emerge in controlled testing environments. 

\begin{table*}[htbp]
\centering
\small
\vspace{2mm}
\caption{Institutional actors involved in AI auditing and the levels of access typically available to them. Different actors contribute complementary forms of oversight, ranging from internal technical audits to independent public scrutiny.}

\begin{tabular}{p{5.75cm}p{1.25cm}p{2.55cm}p{6.5cm}}
\toprule
\textbf{Auditor Type} & \textbf{Access Level} & \textbf{Role} & \textbf{Example Audit Tasks} \\
\midrule

Internal Organizational Teams~\citep{raji2020closing, karnik2024embodied}
& White-box 
& Internal oversight and risk management 
& Audit training pipelines and datasets; conduct internal safety evaluations and red-teaming \\\hline

Trusted Third-Party Auditors~\citep{Sandvig2014AuditingA}
& Grey-box 
& Independent technical evaluation 
& Assess model documentation and evaluation protocols; test for safety or fairness issues under controlled access \\\hline

Regulators and Government Agencies~\citep{Tabassi2023-rl}
& Grey-box / White-box 
& Regulatory oversight and compliance checks 
& Verify safety adherence, privacy, or consumer protection requirements; review system documentation or logs \\\hline

External Researchers and Civil Society~\citep{Sandvig2014AuditingA, Metaxa2021}
& Black-box 
& Independent scrutiny of deployed systems 
& Probe systems by testing inputs and outputs; detect biases, unsafe behaviors, or unintended model responses \\\hline

User Audits and Participatory Governance~\citep{Delgado2023,veale2018, Stanford2026IndustryWideForum, parthasarathy2024participatoryapproachesaidevelopment}
& Black-box 
& Real-world feedback and incident reporting 
& Surface unexpected failures or harmful outputs during everyday system use; report incidents and emerging risks \\

\bottomrule
\end{tabular}
\label{tab:audit_roles}
\end{table*}

\subsection{Targets for AI Auditing}
Auditing AI systems requires examining multiple aspects of system behavior and impact. A recent report from NIST highlights six categories for post-deployment monitoring~\citep{Rao2026-zq}, which are useful for categorizing AI auditing while allowing interactions between categories (e.g., human-AI interactions can cause safety failures).

\paragraph{}\hspace{-5.5mm}\textbf{Functional Audits.} 
Functional auditing can be seen as an extension of benchmark evaluations, but focused on whether an AI system continues to perform reliably as intended after deployment. This includes evaluating model performance under real-world conditions, identifying degradation due to distribution shifts, and detecting unanticipated inputs or use cases. 

\paragraph{}\hspace{-5.5mm}\textbf{Operational Audits.} Operational auditing focuses on the broader system, including the infrastructure that underlies the AI deployment (e.g., data collection pipelines, logging mechanisms, and service reliability (uptime)). Failures in AI systems aren't necessarily due to the model; they can also come from other parts of the system or from the integration of different subsystems. Operational audits can be seen as evaluating system-level consistency and reliability.

\paragraph{}\hspace{-5.5mm}\textbf{Human-AI Interaction Audits.} 
In many cases, AI systems operate as part of human-AI teams, where humans interpret and act on model outputs. Auditing these interactions requires evaluating how users understand, trust, and respond to model recommendations--inspecting the feedback loops and the potential for automation bias or misuse. This is particularly challenging as there are no good benchmarks, and it is unclear where to start for evaluation.

\paragraph{}\hspace{-5.5mm}\textbf{Safety and Security Audits.} 
Safety and security auditing focuses on whether AI systems are resilient to misuse, adversarial inputs, and malicious attacks. This includes evaluating robustness to adversarial manipulation and assessing guardrails against deceptive or unsafe system behavior. These audits are particularly important for systems deployed in high-impact or adversarial environments. Recent work has also explored safety protocols designed to prevent models from intentionally subverting safeguards. For example, \citet{greenblatt2024control} proposes ``control evaluations,'' a methodology for evaluating safety protocols by simulating adversarial models that attempt to bypass monitoring and auditing mechanisms. However, in such settings, when we use a separate AI model as the auditor, it is natural to ask \textit{who audits the auditor?} For this reason, we benefit from considering auditing as a statistical framework, which further motivates our proposed perspective of treating auditing as a risk-control framework.

\paragraph{}\hspace{-5.5mm}\textbf{Compliance Audits.} 
Compliance auditing assesses whether AI systems adhere to relevant laws, regulations, and organizational policies. This may include verifying that systems meet requirements related to fairness, privacy, safety standards, or domain-specific regulatory obligations. As regulatory frameworks for AI continue to evolve, compliance audits play an important role in ensuring that organizations deploy AI systems responsibly.

\paragraph{}\hspace{-5.5mm}\textbf{Large-scale Impact Audits.} 
Large-scale impact auditing considers the broader effects of AI systems once deployed. This includes identifying unintended harms, discriminatory outcomes, or systemic risks that may emerge across populations or institutions. This is especially important in the current AI climate, with frontier AI companies deploying models that garner millions of monthly users\footnote{There were a reported 18.9 million monthly active Claude AI users in early 2026.} from people across the world. Such audits often require longitudinal analysis and interdisciplinary approaches, as the societal consequences of AI systems may evolve and vary across contexts.

\section{Blue-Sky Vision and Challenges}
\noindent {\bf Position Statement:} We argue for the development of principled auditing tools that assess AI systems in the wild and provide feedback mechanisms to maintain compliance with desired properties. We view the auditing of AI systems as a statistical problem of \textit{monitoring constraint violations under uncertainty}. Modern AI systems are deployed in dynamic environments where data distributions shift, user behavior evolves, and ground truth labels may be delayed or unavailable. Consequently, auditing cannot rely solely on deterministic evaluation metrics or static benchmarks; instead, it must quantify the \textit{risk} that deployed systems violate important constraints such as fairness, safety, or regulatory compliance.

\begin{figure}
    \begin{subfigure}{0.5\linewidth}
        \centering
        \includegraphics[width=\linewidth]{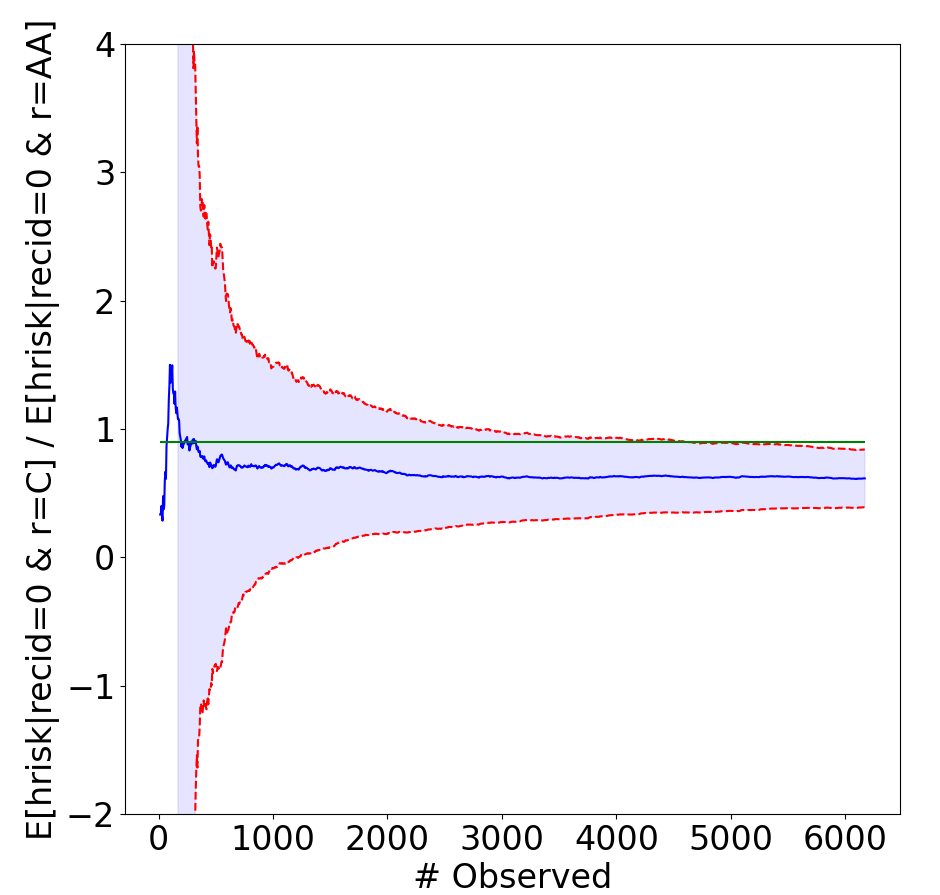}
        \label{fig:casestudy:compas:propublica}
    \end{subfigure}%
    \begin{subfigure}{0.5\linewidth}
        \centering
        \includegraphics[width=\linewidth]{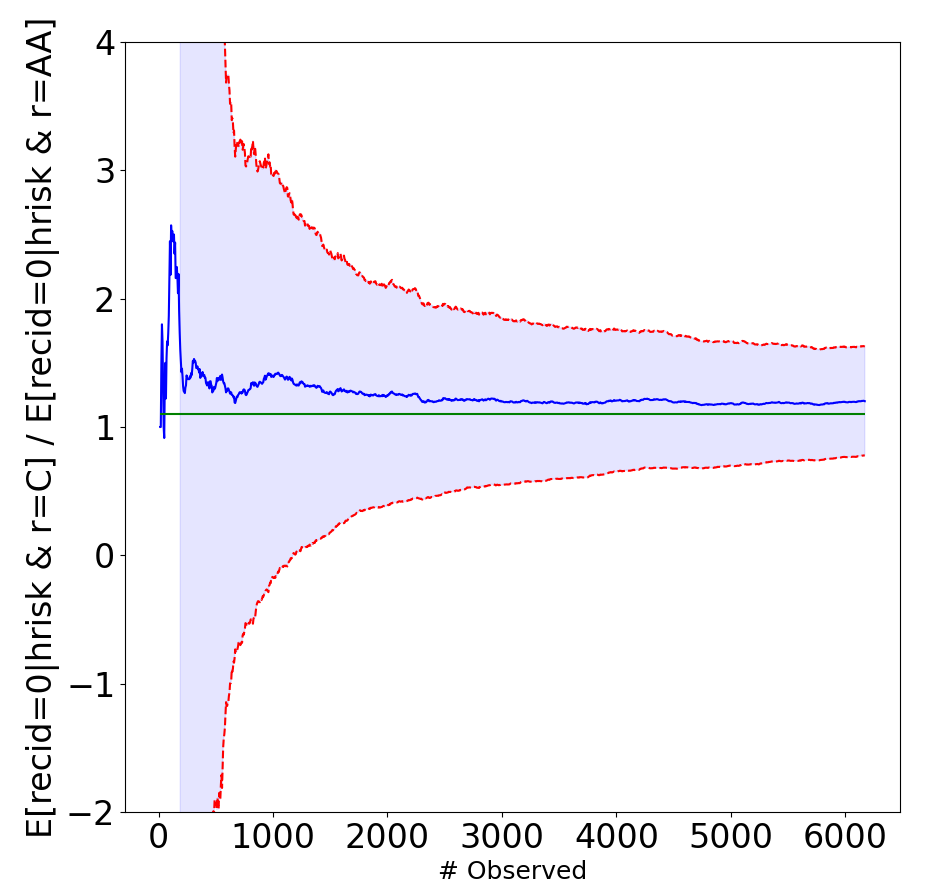}
        \label{fig:casestudy:compas:northpointe}
    \end{subfigure}

    \caption{COMPAS dataset case study (Image source:~\citep{maneriker2023}).\\(a) Analysis done by ProPublica using False Positive Rate Bias specification demonstrating a fairness violation. (b) Analysis done by Northpointe using the False Discovery Rate.}
    \Description{Two plots from a COMPAS fairness case study comparing fairness-auditing outcomes for different bias specifications. Panel (a) shows the ProPublica ``Sample A'' false-positive rate analysis; the estimated fairness metric converges below the 0.9 threshold, indicating the model violates the required fairness guarantee with high confidence. Panel (b) shows the Northpointe “Sample B” false discovery rate analysis; although the point estimate appears near the target threshold, the confidence bounds do not converge sufficiently to verify the fairness claim. Vertical markers indicate stopping points for the auditing procedure, and horizontal reference lines mark the fairness threshold.}
    \vspace{-1mm}
    \label{fig:compas_avoir}
\end{figure}

Under this perspective, auditing becomes a problem of continuously monitoring whether deployed systems satisfy specified constraints as they interact with real-world environments. Because observations of system behavior may be noisy, incomplete, or influenced by human–AI interactions and surrounding infrastructure, auditing must reason about uncertainty in both the data and evaluation processes. This framing highlights several core challenges for AI auditing: specifying measurable constraints, estimating system behavior under uncertainty, and designing auditing mechanisms that operate continuously as systems evolve over time.
\begin{figure*}
    \centering
    \includegraphics[width=\linewidth]{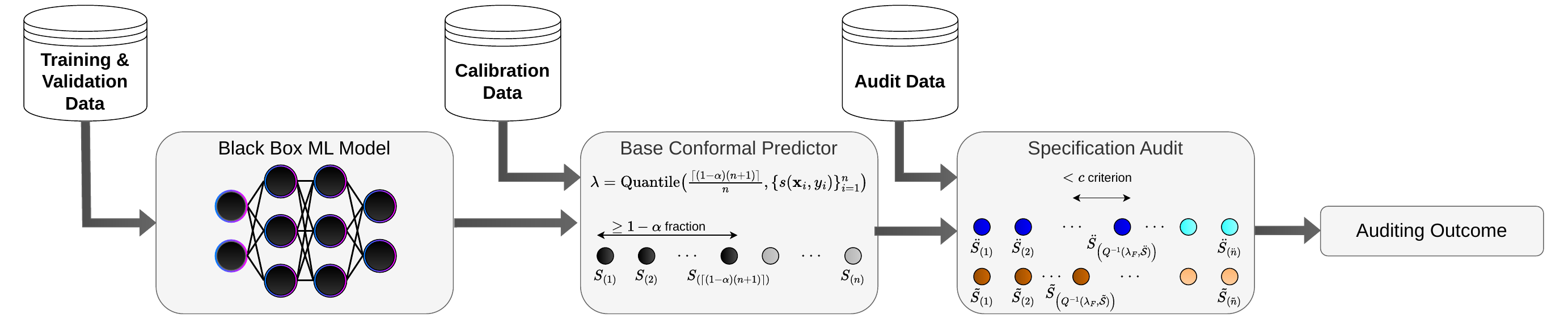}
    \caption{End-to-end Conformal Fairness pipeline. Once a classification model is trained, a conformal predictor is constructed and later audited using a held-out auditing set that is \textit{exchangeable} with the original calibration data. (Image source: \citep{vadlamani2025a})}
    \Description{Pipeline figure showing three stages. The first is training a classification model with training/validation data. The second is constructing a conformal predictor using separate calibration data. The third is for auditing the conformal predictor using separate auditing data.}
    \label{fig:cf_audit}
\end{figure*}
\paragraph{}\hspace{-5.5mm}\textbf{Lack of Trusted Specifications.} 
Algorithmic fairness in machine learning~\citep{barocas2023fairness}~has been a sandbox for developing auditing tools~\citep{ghosh2021justicia,maneriker2023, yan2022active}, where notions of fairness have been specified in regulations (e.g., Four-Fifths Rule ~\citep{eeoc1979}). For example, figure~\ref{fig:compas_avoir} provides an example of fairness auditing as it pertains to the COMPAS recidivism dispute, where AVOIR~\citep{maneriker2023} is used to validate both ProPublica's and NorthPointe's claims about the recidivism system's fairness, which depended on the specific fairness specification. However, the notion of safety in AI systems is less well-defined, with many safety and alignment works operating under the perspective that ``we know unsafe behavior when we see it'' \citep{christiano2017,ouyang2022traininglanguagemodelsfollow}. One of the key challenges and initiatives the community is starting to pursue is defining and specifying safety more precisely as something we can monitor and audit against. Individual domains, such as environmental and financial law, have defined their own specifications for specific use cases, suggesting that specifications need to be very siloed to make them quantifiable or evaluable, but also suggesting that we can pull from these other domains to come up with a way to define socio-technical specifications for AI safety.

\paragraph{}\hspace{-5.5mm}\textbf{Functional Auditing Challenges.} 
Auditing models deployed in real-world, dynamic environments introduces new challenges, including drift, missing ground truth, and unexpected user behavior, which complicate evaluation~\citep{ojewale2025towards}. A risk-based perspective would enable probabilistic measurement of constraints such as fairness, safety, and compliance, providing a statistical foundation for auditing. Recent works have explored methods to incorporate fairness constraints into uncertainty quantification frameworks~\citep{romano2020malice, vadlamani2025a, srinivasan2026fedcf}. Conformal Fairness (CF) is one such work that provides a method for performing fairness auditing in settings where the i.i.d. assumption does not hold post-deployment. Figure~\ref{fig:cf_audit} illustrates the end-to-end pipeline for auditing post-deployment. A similar risk-based perspective can be applied to safety, with clearly defined specifications that accommodate context and uncertainty. While these methods use Conformal Prediction~\citep{vovk2005algorithmic} as the underlying statistical framework, other statistical frameworks can be used to give guarantees on model performance~\citep{angelopoulos2023predictionpoweredinference, bashari2025syntheticpoweredpredictiveinference}.

\paragraph{}\hspace{-5.5mm}\textbf{Operational, Data, and Compliance Challenges.} 
AI infrastructure, data pipelines, and provenance tracking are fragmented in practice, complicating auditing~\citep{ojewale2025towards}. In addition to evaluating and auditing the learning models, understanding the data used to train them is just as crucial~\citep{sambasivan2021everyone}, but is less well explored. There has been some work on auditing data membership~\citep{Huang_2024}. There are also ways to characterize properties of data that determine the effectiveness, accuracy, and scalability of machine learning models, including the \textbf{5 Vs of big data}--volume, velocity, variety, veracity, and value~\citep{Demchenko2013}. Data audits often examine these dimensions to assess dataset quality and identify issues that may affect model performance. Data audits are further complicated by federated deployments, data sovereignty, and heterogeneous data types and modalities~\citep{chang2024efficient}. A similar risk-based auditing perspective to Conformal Fairness can be applied in federated settings~\citep{srinivasan2026fedcf}, but the area remains ripe for exploration. Compliance audits must navigate evolving policy landscapes and ensure consistency across pre- and post-deployment phases. Addressing these challenges requires socio-technical solutions that integrate technical, legal, and organizational perspectives.

\paragraph{}\hspace{-5.5mm}\textbf{Human-AI Teams and Socio-Technical Challenges.} 
Auditing models with humans in the loop, as users of AI (Human-AI teams) or as auditors, obfuscates model behavior with human intent~\citep{chen2023human}, thereby undermining the reliability of the auditing results. The auditing problem is analogous to partial-observation problems in reinforcement learning and classical control theory \citep{monahan1982state}, in which the true state (the audit results) must be distilled from raw, noisy observations. However, this is challenged by the unintelligible nature of human behavior and responses to information from ML models~\citep{small2023helpful, cresswell2025conformal}. To redress, we need to develop approaches that are robust to bias and uncertainty stemming from human behavior, draw on insights from the partial observability literature, and balance the socio-technical with humans-in-the-loop, while maintaining data privacy and human safety. Furthermore, the epistemic uncertainty of AI agents, combined with the cognitive uncertainty in human decision-making, suggests that auditing should shift toward quantifying the risk associated with the \textbf{joint behavior} of human-AI teams, enabling a more reliable assessment of system compliance and performance. The increasing prevalence of AI agents for a broad spectrum of tasks necessitates further consideration.

\paragraph{}\hspace{-5.5mm}\textbf{Challenges with Modern AI Agents in Human-AI Teams.} 
In a recent case study, many users recognized the potential value of agents for low-risk, repetitive tasks, but were skeptical of using AI agents for complex medical and financial decisions, particularly when errors could not be corrected~\citep{Stanford2026IndustryWideForum}. Trust increased when AI agents offered transparency, user control, and step‑by‑step explanations. For silos, including health or finance, there was strong support for agents requiring explicit consent before completing tasks involving sensitive data. Thus, to increase user trust, a risk-sensitive approach to auditing will be essential for providing concrete guarantees of model performance, particularly in high-risk scenarios. Similarly, closing the development loop by incorporating audit feedback will instill trust in a continually evolving environment. Lastly, \citep{Stanford2026IndustryWideForum} elucidates the need for more user case studies to develop concrete audit specifications for agentic models. 

\paragraph{}\hspace{-5.5mm}\textbf{Risk-Based Perspective on Constraint Violations} 
Many auditing tasks can be interpreted as verifying whether deployed systems satisfy constraints that capture desirable properties such as fairness, safety, reliability, or regulatory compliance. In practice, however, these properties cannot typically be evaluated deterministically due to noisy observations, incomplete data, and evolving deployment environments. As a result, auditing must reason about the \textit{risk} that a system violates a given constraint. Under this perspective, auditing becomes the task of estimating and monitoring the likelihood of constraint violations as systems interact with real-world environments. This framing naturally emphasizes uncertainty-aware auditing methods and motivates approaches that continuously monitor system behavior and trigger intervention when risks exceed acceptable thresholds. Recent work on uncertainty-aware evaluation methods, such as conformal approaches to fairness auditing~\citep{vadlamani2025a, srinivasan2026fedcf}, illustrates one possible direction for operationalizing this perspective. Broadly, risk-control methods for sequential decision-making (cf. \citep{angelopoulos2024conformal, xu2024active, ramdas2022admissibleanytimevalidsequentialinference, ramdas2023game, prinster2025watch}) provide a relevant basis for encoding auditing specifications and for maintaining rigorous guarantees. This perspective provides a method for quantifying uncertainty in the auditing process, which we can then use to inform the feedback loop for model development. This view aligns with causal and mechanism-aware anomaly detection, which treats failures as violations of stable system invariants rather than distributional shifts~\citep{peters2016causal,arjovsky2020invariantriskminimization}.

\section{Concluding Remarks}
\paragraph{}\hspace{-5.5mm}\textbf{What would success look like for AI auditing?} Under this risk-based perspective, we would be able to quantify the uncertainty or the ``risk'' associated with an audit criterion, which is then used to inform subsequent model updates. If the information used to perform subsequent development results in a model that passes the audit (or controls the ``risk''), we consider the audit a success.

Advances across multiple disciplines--including law, governance, management, economics, and health sciences, as well as computer science, engineering, signal processing, mathematics, and statistics--will be necessary to address the challenges of auditing modern AI systems. These challenges span the specification of measurable constraints (e.g., fairness, safety, and compliance), the development of mechanisms for monitoring deployed systems under uncertainty, and the design of auditing processes that operate throughout the lifecycle of an AI system. Comparable auditing practices already exist in sectors such as finance, healthcare, pharmaceuticals, and environmental regulation, where systems are monitored continuously and assessed against evolving regulatory and safety standards. An important direction for future work is therefore to examine how principles from these established auditing frameworks can inform the development of uncertainty-aware auditing mechanisms for modern AI systems across the AI lifecycle.

\begin{acks}
The authors acknowledge support from National Science Foundation (NSF) grant \#2112471 (AI-EDGE). The authors' views and findings do not necessarily reflect those of the funding agencies.
\end{acks}

\bibliographystyle{ACM-Reference-Format}
\balance
\bibliography{references}

\end{document}